\documentclass[a4paper]{article}

\usepackage{makeidx}
\usepackage{amsfonts}
\usepackage{amsmath}
\usepackage[dvips]{graphicx}
\usepackage{url}
\usepackage{multirow}
\usepackage{setspace}
\usepackage{verbatim}
\usepackage{algorithm}
\usepackage{algorithmicx}
\usepackage{algpseudocode}
\usepackage{authblk}

\usepackage{subfig}
\usepackage{multicol}
\usepackage{array}
\usepackage{fancyhdr}

\newcommand\blfootnote[1]{%
  \begingroup
  \renewcommand\thefootnote{}\footnote{#1}%
  \addtocounter{footnote}{-1}%
  \endgroup
}

\begin{document}

%\mainmatter  % start of an individual contribution

\title{Blocked All-Pairs Shortest Paths Algorithm on Intel Xeon Phi KNL Processor: A Case Study}

%\titlerunning{Blocked All-Pairs Shortest Paths Algorithm on Intel Xeon Phi KNL Processor: A Case Study}

\author[1]{Enzo Rucci\thanks{erucci@lidi.info.unlp.edu.ar}}
\author[1]{Armando De Giusti}
\author[2]{Marcelo Naiouf}
\affil[1]{III-LIDI, CONICET, Facultad de Inform\'atica, Universidad Nacional de La Plata}
\affil[2]{III-LIDI, Facultad de Inform\'atica, Universidad Nacional de La Plata}

\maketitle

%-------------------------------------------------------------------------------

%Nowadays accelerators have consolidated in HPC community as a way of improving performance while keeping power efficiency. Recently, Intel has released the second generation of its Xeon Phi family, known as KNL.

\begin{abstract}
Manycores are consolidating in HPC community as a way of improving performance while keeping power efficiency. 
Knights Landing is the recently released second generation of Intel Xeon Phi architecture. While optimizing applications on CPUs, GPUs and first Xeon Phi's has been largely studied in the last years, the new features in Knights Landing processors require the revision of programming and optimization techniques for these devices. 
In this work, we selected the Floyd-Warshall algorithm as a representative case study of graph and memory-bound applications. Starting from the default serial version, we show how data, thread and compiler level optimizations help the parallel implementation to reach 338 GFLOPS.

\end{abstract}

\blfootnote{The final publication is available at Springer via \url{https://doi.org/10.1007/978-3-319-75214-3_5}}

\section{Introduction}
\label{sec:intro}

% Consolidación de aceleradores en HPC. Top500 and green500. uso de multicore con gpus y mic.
The power consumption problem represents one of the major obstacles for Exascale systems design. As a
consequence, the scientific community is searching for different ways to improve power efficiency of High Performance Computing (HPC) systems~\cite{Giles2014}. One recent trend to increase compute power and, at the same time, limit power consumption of these systems lies in adding accelerators, like NVIDIA/AMD graphic processing units (GPUs), or Intel Many Integrated Core (MIC) co-processors. These manycore devices are capable of achieving better FLOPS/Watt ratios than traditional CPUs. For example, the number of Top500~\cite{Top500} systems using accelerator technology grew from 54 in June 2013 to 91 in June 2017. In the same period, the number of systems based on accelerators increased from 55 to 90 on the Green500 list~\cite{Green500}.

% recientemente intel presentó KNL. diferencias con KNC. importancia de programación y otpomización.
Recently, Intel has presented the second generation of its MIC architecture (branded Xeon Phi), codenamed Knigths Landing (KNL).  Among the main differences of KNL regarding its predecessor Knights Corner (KNC), we can find the incorporation of AVX-512 extensions, a remarkable number of vector units increment, a new on-package high-bandwidth memory (HBM) and the ability to operate as a standalone processor. Even though optimizing applications on CPUs, GPUs and KNC Xeon Phi's has been largely studied in the last years, accelerating applications on KNL processors is still a pending task due to its recent commercialization. In that sense, the new features in KNL processors require the revision of programming and optimization techniques for these devices. 

In this work, we selected the Floyd-Warshall (FW) algorithm as a representative case study of graph and memory-bound applications. This algorithm finds the shortest paths between all pairs of vertices in a graph and occurs in domains of communication networking~\cite{Floyd_networking}, traffic routing~\cite{floyd_traffic}, bioinformatics~\cite{floyd_bioinformatics}, among others. FW is both computationally and spatially expensive since it requires $O(n^3)$ operations and $O(n^2)$ memory space, where $n$ is the number
of vertices in a graph. Starting from the default serial version, we show how data, thread and compiler level optimizations help the parallel implementation to reach 338 GFLOPS.

% Contenido del paper 
The rest of the present paper is organized as follows. Section~\ref{sec:xeon_phi} briefly introduces the Intel Xeon Phi KNL architecture while Section~\ref{sec:floyd} presents the FW algorithm. Section~\ref{sec:implementation} describes our implementation. In Section~\ref{sec:results} we analyze performance results while Section~\ref{sec:related_works} discusses related works. Finally, Section~\ref{sec:conclusion} outlines conclusions and future lines of work.

\section{Intel Xeon Phi Knights Landing}
\label{sec:xeon_phi}

%In 2012, Intel launches the KNC generation which mainly features up to 61 x86 pentium cores with extended vector units (512-bit) and simultaneous multithreading (4 threads per core). 
KNL is the second generation of the Intel Xeon Phi family and the first capable of operating as a standalone processor. The KNL architecture is based on a set of \emph{Tiles} (up to 36) interconnected by a 2D mesh. Each Tile includes 2 cores based on the out-of-order Intel's Atom micro-architecture (4 threads per core), 2 Vector Processing Units (VPUs) and a shared L2 cache of 1 MB. These VPUs not only implement the new 512-bit AVX-512 ISA but they are also compatible with prior vector ISA's such as SSE\emph{x} and AVX\emph{x}. 
AVX-512 provides 512-bit SIMD support, 32 logical registers, 8 new mask registers for vector predication, and gather and scatter instructions to support loading and storing sparse data. As each
AVX-512 instruction can perform 8 double-precision (DP) operations
(8 FLOPS) or 16 single-precision (SP)  operations (16 FLOPS), the peak performance is over 1.5 TFLOPS in DP and 3 TFLOPS in SP, more than two times higher than that of the KNC. It is also more energy efficient than its predecessor~\cite{KNLbook}.

Other significant feature of the KNL architecture is the inclusion of an in-package HBM called MCDRAM. This special memory offers 3 operating modes: \emph{Cache}, \emph{Flat} and \emph{Hybrid}. In Cache mode, the MCDRAM is used like an L3 cache, caching data from the DDR4 level
of memory. Even though application code remains unchanged, the MCDRAM can suffer lower performance rates.  In Flat mode, the MCDRAM has a physical addressable space offering the highest bandwidth
and lowest latency. However, software modifications may be required in order to use both the DDR and the MCDRAM in the same application. Finally, in the \emph{Hybrid mode}, HBM is divided in two parts: one part in \emph{Cache mode} and one in \emph{Flat mode}~\cite{knl_best_practice_guide}.

%\begin{figure}[tb]
%	\begin{centering}
%	\includegraphics[width=0.9\columnwidth]{figs/phi-knl.eps}
%	\par
%	\end{centering}
%	\caption{\label{fig:phifig} Xeon Phi KNL architecture.}
%\end{figure}

From a software perspective, KNL supports  parallel programming models used traditionally on HPC systems such as OpenMP or MPI. This fact represents a strength of this platform since it simplifies code development and improves portability over other alternatives based on accelerator specific programming languages such as CUDA or OpenCL.
 However, to achieve high performance, programmers should attend to:
\begin{itemize}
\item the efficient exploitation of the memory hierarchy, especially
when handling large datasets, and
\item how to structure the computations to take advantage of the VPUs.
\end{itemize}
Automatic vectorization is obviously the easiest programming way to exploit VPUs. However, in most cases the compiler is unable to generate SIMD binary code since it can not detect free data dependences into loops. In that sense, SIMD instructions are supported in KNL processors through the use of guided compilation or hand-tuned codification with intrinsic instructions~\cite{KNLbook}.  On one hand, in guided vectorization,  the programmer indicates the compiler (through the insertion of tags) which loops are independent and their memory pattern access. In this way, the compiler is able to generate SIMD binary code preserving the program portability. On the other hand, intrinsic vectorization usually involves rewriting most of the corresponding algorithm. The programmer gains in control at the cost of losing portability. Moreover, this approach also suggests the inhibition of other compiler loop-level optimizations. Nevertheless, it is the only way to exploit parallelism in some applications with no regular access patterns or loop data dependencies which can be hidden by recomputing techniques~\cite{Culler97}.

%However, minimal programming efforts such as the introduction of some directives to inform the compiler about pointer disambiguation or data alignment data dependencies usually provide poor performance rates. In fact, \textit{guided auto-vectorization} is not able to achieve the best performance in most cases and programmers usually need to make an extra effort to hand-tune the codes to exploit SIMD capabilities. Indeed, intrinsics are currently the only option for complex applications which suffer from data dependencies or irregular access patterns that can be hidden using specific code transformations. Unfortunately, improving performance comes at the expense of losing cross-platform portability. %Most processor families, even from the same vendor, have non-compatible intrinsics which support different SIMD instruction sets. As a consequence, code developers need to write many code branches, thus increasing maintenance needs.

\section{Floyd-Warshall Algorithm}
\label{sec:floyd}

\begin{figure}[b]
    \centering
    \begin{minipage}{0.53\textwidth}
        \centering
        \includegraphics[width=1\textwidth]{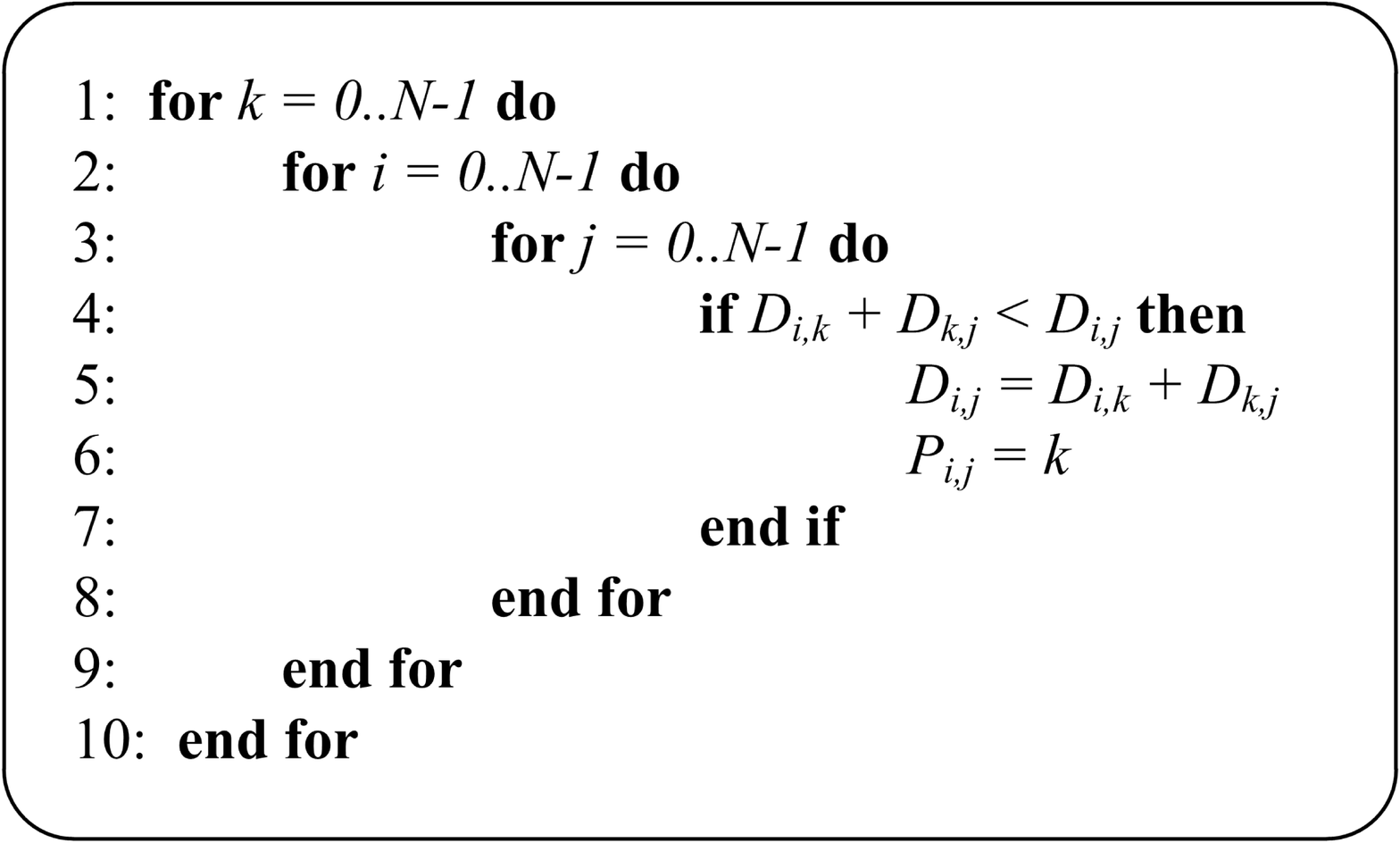} % second figure itself
        \caption{\label{fig:floyd_naive} Naive Floyd-Warshall Algorithm}
    \end{minipage}\hfill
    \begin{minipage}{0.43\textwidth}
        \centering
        \includegraphics[width=1\textwidth]{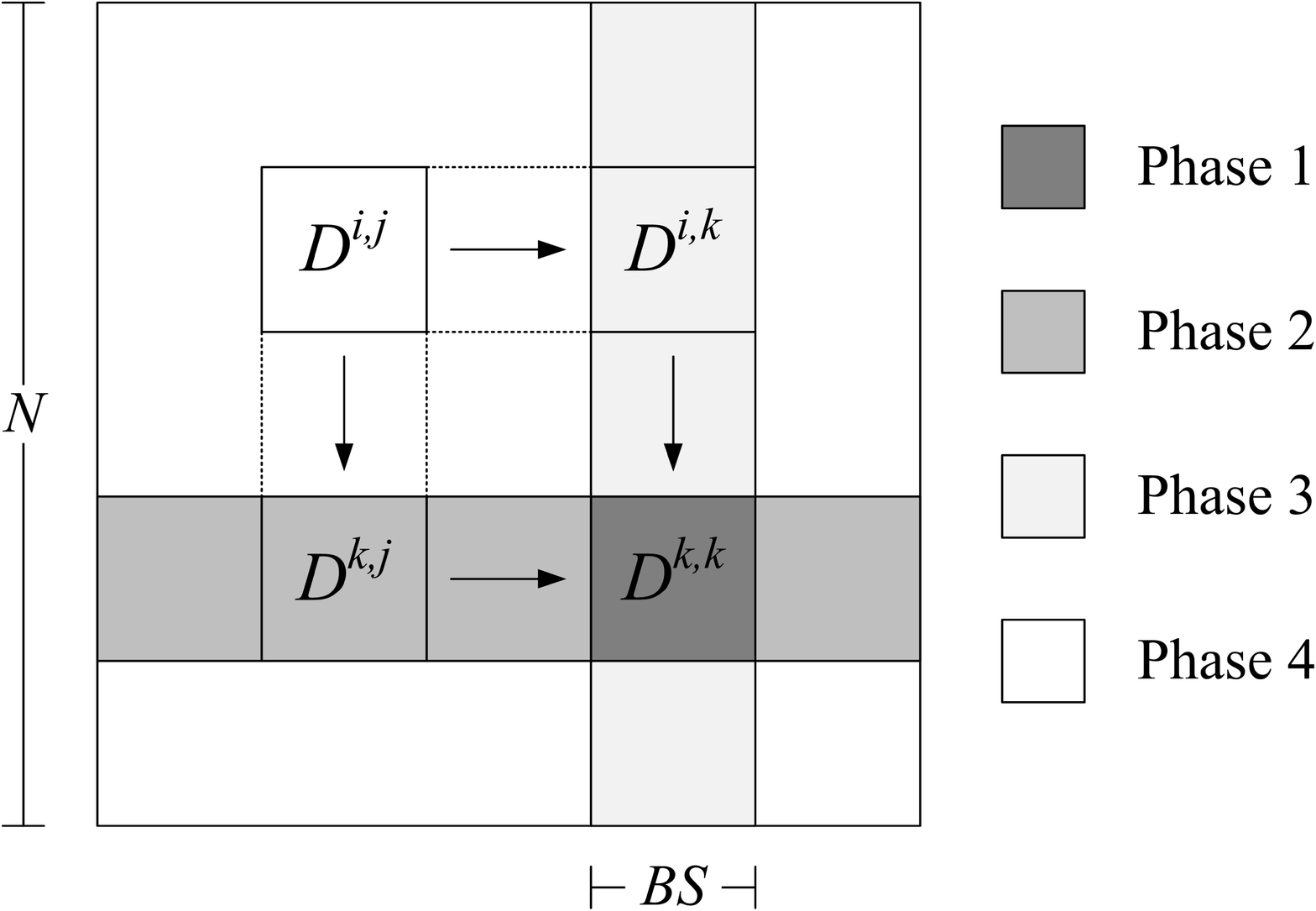} % second figure itself
        \caption{\label{fig:floyd_blocked} Schematic representation of the blocked Floyd-Warshall Algorithm}
    \end{minipage}
\end{figure}

The FW algorithm uses a dynamic programming approach to compute the all-pairs shortest-paths
problem on a directed graph~\cite{Floyd,Warshall}. 
This algorithm takes as input a $N \times N$ distance matrix $D$, where $D_{i,j}$ is initialized with the original distance from node \emph{i} to node \emph{j}. FW runs for $N$ iterations and at \emph{k}-th iteration it evaluates all the possible paths between each pair of vertices from \emph{i} to \emph{j} through the intermediate vertex \emph{k}. As a result, FW produces an updated matrix $D$, where $D_{i,j}$ now contains the shortest distance between nodes \emph{i} and \emph{j}. Besides, an additional matrix $P$ is generated when the reconstruction of the shortest path is required. $P_{i,j}$ contains the most recently added intermediate node between \emph{i} and \emph{j}. Figure~\ref{fig:floyd_naive} exhibits the naive FW algorithm.

%\input{alg1}

% 2003 - a blocked all pair...
%2005 - optimizing graph algorithms ...
% 2006 - hardware-software 
% 2005 - OPTIMIZING ALL PAIRS

\section{Implementation}
\label{sec:implementation}

In this section, we address the optimizations performed on the Intel Xeon Phi KNL processor. First of all, we developed a serial implementation following the naive version described in Figure~\ref{fig:floyd_naive}, as this implementation will work as baseline. Next, we optimized the serial version considering data locality and data level parallelism. Finally, we introduced thread level parallelism exploiting the OpenMP programming model to obtain a multi-threaded implementation.

\subsection{Data Locality}
\label{subsec:data_locality}

\begin{figure}[t]
\begin{centering}
\includegraphics[width=0.9\columnwidth]{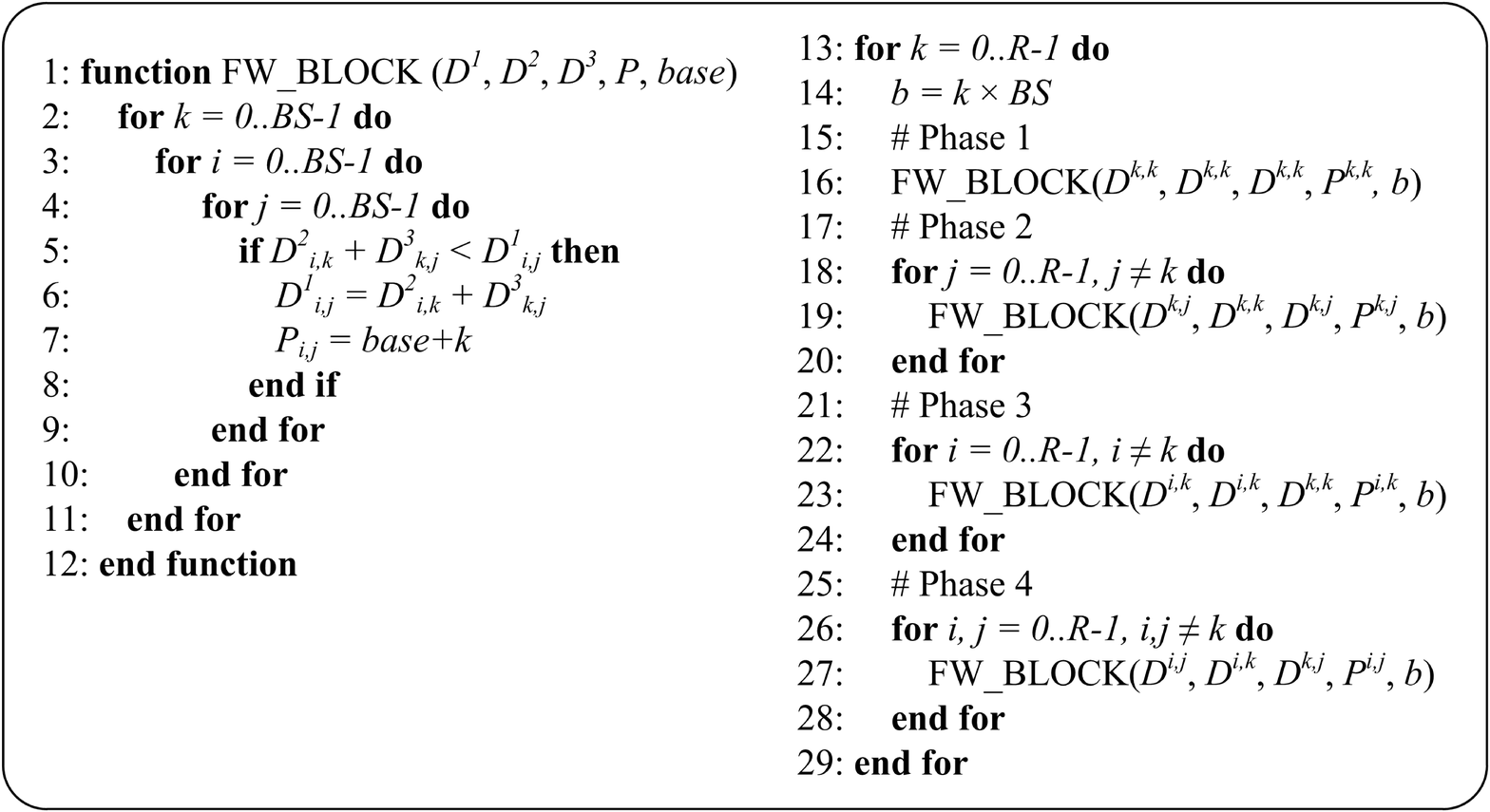}
\par\end{centering}
\caption{\label{fig:floyd_blocked_alg} Blocked Floyd-Warshall algorithm.}
\end{figure}

To improve data locality, the FW algorithm can be blocked~\cite{floyd_blocked2}. Unfortunately, the three loops can not be interchanged in free manner due to the data dependencies from one iteration to the next in the \emph{k}-loop (just \emph{i} and \emph{j} loops can be done in any order). However, under certain conditions, the \emph{k}-loop can be put inside the \emph{i}-loop and \emph{j}-loop, making blocking possible. 
The distance matrix  $D$ is partitioned into blocks of size $BS \times BS$, so that there are $(N/BS)^2$ blocks. The computations involve $R=N/BS$ rounds and each round is divided into four phases based on the data dependency among the blocks:
\begin{enumerate}
\item Update the block \emph{k,k} ($D^{k,k}$) because it is self-dependent.
\item Update the remaining blocks of the \emph{k}-th row because each of these blocks depends on itself and the previously computed $D^{k,k}$.
\item Update the remaining blocks of the \emph{k}-th column  because each of these blocks depends on itself and the previously computed $D^{k,k}$.
\item Update the rest of the matrix blocks as each of them depends on the \emph{k}-th block of its row and the \emph{k}-th block of its column.
\end{enumerate}
In this way, we satisfy all dependencies from this algorithm. Figure~\ref{fig:floyd_blocked} shows a schematic representation of a round computation and the data dependences among the blocks while Figure~\ref{fig:floyd_blocked_alg} presents the corresponding pseudo-code.

\subsection{Data Level Parallelism}
\label{subsec:data_level_parallelism}

\begin{figure}[t]
    \centering
    \begin{minipage}{0.43\textwidth}
        \centering
        \includegraphics[width=1\textwidth]{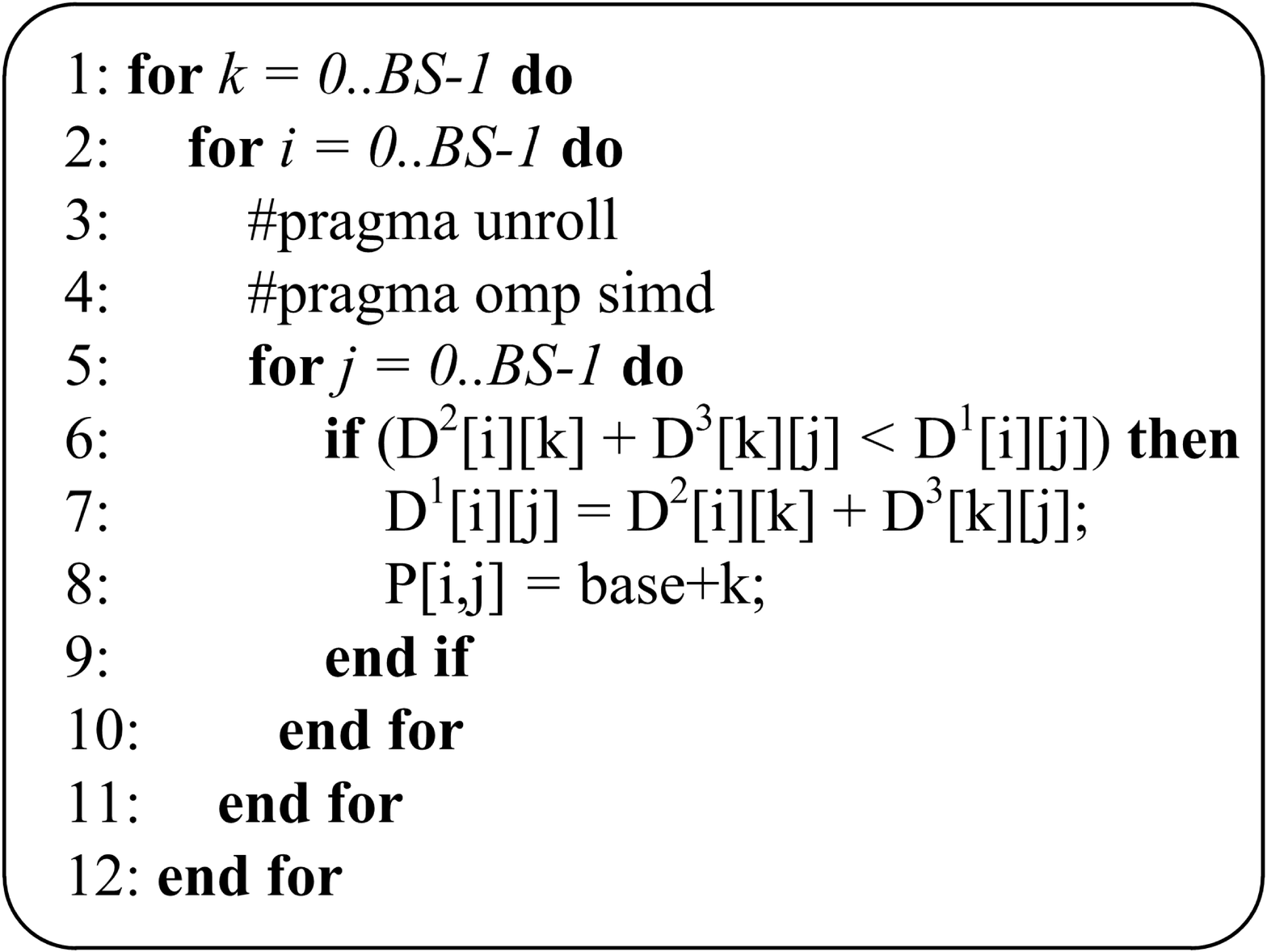} % second figure itself
        \caption{\label{fig:fw_block_guided} Pseudo-code for FW\_BLOCK implementation using guided vectorization}
    \end{minipage}\hfill
    \begin{minipage}{0.53\textwidth}
        \centering
        \includegraphics[width=1\textwidth]{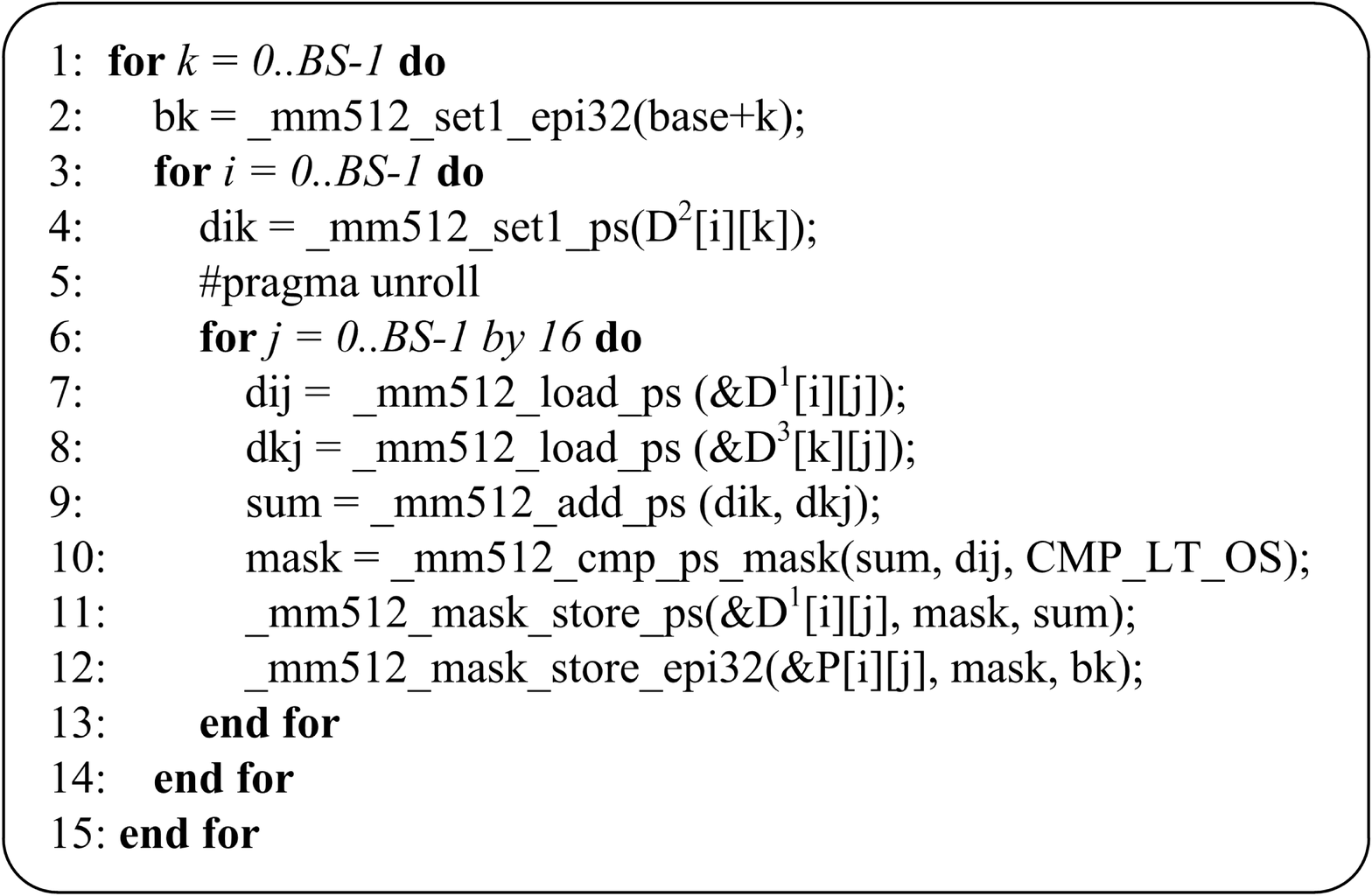} % second figure itself
        \caption{\label{fig:fw_block_intrinsic} Pseudo-code for FW\_BLOCK implementation using intrinsic vectorization}
    \end{minipage}
\end{figure}

The innermost loop of FW\_BLOCK code block from Figure~\ref{fig:floyd_blocked_alg} is clearly the most computationally expensive part of the algorithm. In that sense, this loop is the best candidate for vectorization. The loop body is composed of an \emph{if} statement that involves one addition, one comparison and (may be) two assign operations. Unfortunately, the compiler detects false dependencies in that loop and is not able to generate SIMD binary code. For that reason, we have explored two SIMD exploitation approaches: (1) guided vectorization through the usage of the OpenMP 4.0 \emph{simd} directive  and (2) intrinsic vectorization employing the AVX-512 extensions. The guided approach simply consists of inserting the \emph{simd} directive to the innermost loop of FW\_BLOCK code block (line 4). On the opposite sense, the intrinsic approach consists of rewriting the entire loop body. Figures~\ref{fig:fw_block_guided} and~\ref{fig:fw_block_intrinsic} show the pseudo-code for FW\_BLOCK implementation using guided and manual vectorization, respectively. In order to accelerate SIMD computation with 512-bit vectors, we have carefully managed the memory allocations so that distance and path matrices are 64-byte aligned. In the guided approach, this also requires adding the \emph{aligned} clause to the \emph{simd} directive.

\subsection{Loop Unrolling}
\label{subsec:loop_unrolling}

Loop unrolling is another optimization technique that helped us to improve the code performance. Fully unrolling the innermost loop of FW\_BLOCK code block was found to work well. Unrolling the \emph{i}-loop of the same code block once was also found to work well.

\subsection{Thread Level Parallelism}
\label{subsec:thread_level_parallelism}

To exploit parallelism across multiple cores, we have implemented a multi-threaded version of FW algorithm based on OpenMP programming model. A \emph{parallel} construct is inserted before the loop of line 13 in Figure~\ref{fig:floyd_blocked_alg} to create a parallel block. To respect data dependencies among the block computations, the work-sharing constructs must be carefully inserted. At each round, phase 1 must be computed before the rest. So a \emph{single} construct is inserted to enclose line 16. Next, phases 2 and 3 must be computed before phase 4. As these blocks are independent among them, a \emph{for} directive is inserted before the loops of lines 18 and 22. Besides, 
a \emph{nowait} clause is added to the phase 2 loop to alleviate the thread idling. Finally, another \emph{for} construct is inserted before the loop of line 26 to distribute the remaining blocks among the threads.

\section{Experimental Results}
\label{sec:results}

\subsection{Experimental Design}
All tests have been performed on an Intel server running CentOS 7.2 equipped with a Xeon Phi 7250 processor 68-core 1.40GHz (4 hw thread per core and 16GB MCDRAM memory) and 48GB main memory. The processor was run in \emph{Flat} memory mode and \emph{Quadrant} cluster mode.

We have used Intel's ICC compiler (version 17.0.1.132) with the \emph{-O3} optimization level. To generate explicit AVX2 and AVX-512 instructions, we employed the \emph{-xAVX2} and \emph{-xMIC-AVX512} flags, respectively. Also, we used the \emph{numactl} utility to exploit MCDRAM memory (no source code modification is required). Besides, different workloads were tested: \emph{N} = \{4096, 8192, 16384, 32768, 65536\}. 

\subsection{Performance Results}
\label{sec:perf-knl}

First, we evaluated the performance improvements of the different optimization techniques applied to the naive serial version, such as blocking (\emph{blocked}), data level parallelism (\emph{simd}, \emph{simd (AVX2)} and \emph{simd (AVX-512)}), aligned access (\emph{aligned}) and loop unrolling (\emph{unrolled}). Table~\ref{tab:performance_serial} shows the execution time (in seconds) of the different serial versions when \emph{N}=4096. As it can be observed, blocking optimization reduces execution time by 5\%. 
%Despite this percentage could be considered as a small performance improvement, the blocking technique will also allow a coarser-grain workload distribution in the parallel implementation.
Regarding the block size, 256 $\times$ 256 was found to work best. In the most memory demanding case of each round (phase 4), four blocks are loaded into the cache (3 distance blocks and 1 path block). The four blocks requires 4 $\times$ 256 $\times$ 256 $\times$ 4 bytes = 1024 KB = 1MB, which is exactly the L2 cache size.

As stated in Section~\ref{subsec:data_level_parallelism}, the compiler is not able to generate SIMD binary code by itself in the blocked version. Adding the corresponding \emph{simd} constructs to the blocked version reduced the execution time from 572.66 to 204.52 seconds, which represents a speedup of 2.8$\times$. However, AVX-512 instructions can perform 16 SP operations at the same time. After inspecting the code at assembly level, we realized that the compiler generates SSE\emph{x} instructions by default. As SSE\emph{x} can perform 4 SP operations at the same time, the 2.8$\times$ speedup has more sense since not all the code can be vectorized. Next, we re-compiled the code including the \emph{-xAVX2} and \emph{-xMIC-AVX512} flags to force the compiler to generate AVX2 and AVX-512 SIMD instructions, respectively. AVX2 extensions accelerated the blocked version by a factor of 5.8$\times$ while AVX-512 instructions achieved an speedup of 15.5$\times$. So, it is clear that this application benefits from larger SIMD width. In relation to the other optimization techniques employed, we have found that the \emph{simd (AVX-512)} implementation runs 1.11$\times$ faster when aligning memory accesses in AVX-512 computations (\emph{aligned}). Additionally, applying the loop unrolling optimization to the \emph{aligned} version led to higher performance, gaining a 1.45$\times$ speedup. In summary, we achieve a 26.3$\times$ speedup over the naive serial version through the combination of the different optimizations described.

\begin{table} [t!]
\centering
\caption{\label{tab:performance_serial} Execution time (in seconds) of the different optimization techniques applied to the naive serial version when \emph{N}=4096.}
\begin{tabular*}{12cm}{@{\extracolsep{\fill}}>{\centering}p{1.2cm}>{\centering}m{1.2cm}>{\centering}m{1.4cm}>{\centering}p{2.3cm}>{\centering}m{2.7cm}>{\centering}m{1.3cm}>{\centering}m{1.4cm}}
\hline 
\noalign{\vskip0.2cm}
\emph{naive} & \emph{blocked} & \emph{simd} & \emph{simd (AVX2)} & \emph{simd (AVX-512)} & \emph{aligned} & \emph{unrolled}\tabularnewline
\hline 
\noalign{\vskip0.2cm}
602.8 & 572.66 & 204.52 & 100.47 & 36.95 & 33.28 & 22.95\tabularnewline
\hline 
\end{tabular*}
\end{table}

\begin{figure}[t!]
\begin{centering}
\includegraphics[width=0.9\columnwidth]{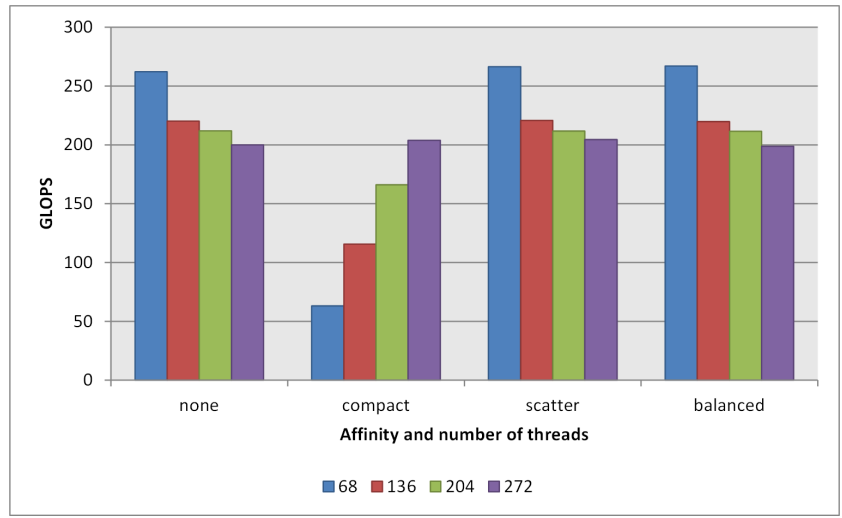}
\par\end{centering}
\caption{\label{fig:performance_threads_affinity}Performance for the different affinity types used varying the number of threads  when \emph{N}=8192.}
\end{figure}

Taking the optimized serial version, we developed a multi-threaded implementation as described in Section~\ref{subsec:thread_level_parallelism}. Figure~\ref{fig:performance_threads_affinity} shows the performance (in terms of GFLOPS) for the different affinity types used varying the number of threads when \emph{N}=8192. As expected, \emph{compact} affinity produced the worst results since it favours using all threads on a core before using other cores. \emph{Scatter} and \emph{balanced} affinities presented similar performances improving the \emph{none} counterpart. As the KNL processor used in this study has all its cores in the same package, \emph{scatter} and \emph{balanced} affinities distribute the threads in the same manner when one thread per core is assigned.
% quizás poner algo del 2D mesh
Regarding the number of threads, using a single thread per core is enough to get maximal performance (except in \emph{compact} affinity). This behavior is opposed to the KNC generation where two or more threads per core where required to achieve high performance. However, it should not be a surprise since the KNL cores were designed to optimize single thread performance including out-of-order pipelines and two VPUs per core.

It is important to remark that, unlike the optimized serial version, the parallel implementation used a smaller block size  since it delivered higher performance. A smaller block size allowed a finer-grain workload distribution and decreased thread idling, especially when the number of threads was larger than the number of blocks in phases 2 and 3. Another reason to decrease block size was that the L2 available space is now shared between the threads in a tile, contrary to the single threaded case. In particular, \emph{BS}=64 was found to work best.

\begin{figure}[t]
\begin{centering}
\includegraphics[width=0.9\columnwidth]{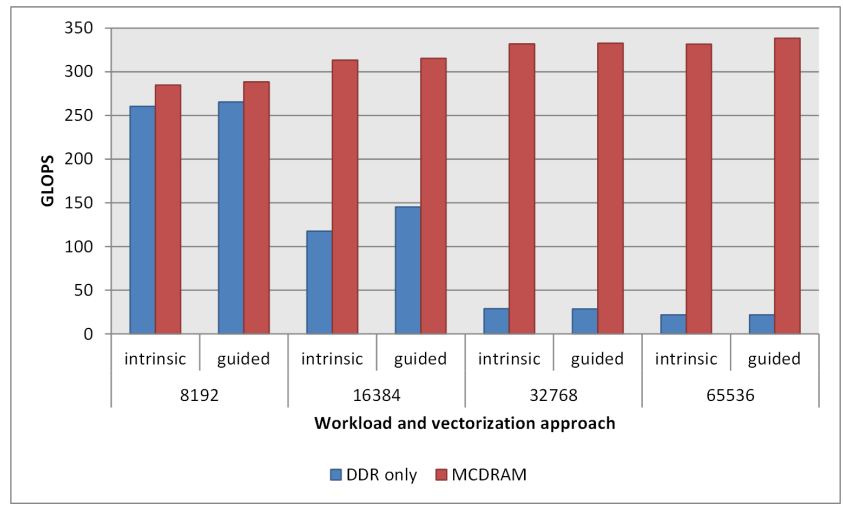}
\par\end{centering}
\caption{\label{fig:performance_n_mcdram}Performance evolution varying workload and the MCDRAM exploitation.}
\end{figure}

Figure~\ref{fig:performance_n_mcdram} illustrates performance evolution varying workload and MCDRAM exploitation for the different vectorization approaches. For small workloads (\emph{N} = 8192), the performance improvement is little ($\sim$1.1$\times$). However, MCDRAM memory presents remarkable speedups for greater workloads, even when the dataset largely exceeds the MCDRAM size (\emph{N} = 655536). In particular, MCDRAM exploitation achieves an average speedup of 9.8$\times$ and a maximum speedup of 15.5$\times$. In this way, we can see how MCDRAM usage is an efficient strategy for bandwidth-sensitive applications.

In relation to the vectorization approach, we can appreciate that guided vectorization leads to slightly better performance than the intrinsic counterpart, running upto 1.03$\times$ faster. The best performances are 330 and 338 GFLOPS for the intrinsic and guided versions, respectively. After analyzing the assembly code, we realized that this difference is caused by the prefetching instructions introduced by the compiler when
guided vectorization is used. Unfortunately, the compiler disables automatic prefetching when code is manually vectorized.

\section{Related Works}
\label{sec:related_works}

Despite its recent commercialization, there are some works that evaluate KNL processors. In that sense, we highlight~\cite{Rosales2016_mcdram} that presents a study of the performance differences observed when using the three MCDRAM configurations available in combination with the three possible memory access or cluster modes. Also, Barnes et al.~\cite{KNL_NERSC} discussed the lessons learned from optimizing a number of different high-performance applications and kernels. Besides, Haidar et al.~\cite{Haidar2016_KNL} proposed and evaluated several optimization techniques for different matrix factorizations methods on many-core systems.

Obtaining high-performance in graph algorithms is usually a difficult task since they tend to suffer from irregular dependencies and large space requirements. Regarding FW algorithm, there are many works proposed to solve the all-pairs shortest paths problem on different harwdare architectures. However, to the best of the authors knowledge, there are no related works with KNL processors.
Han and Kang~\cite{floyd_han} demonstrated that exploiting SSE2 instructions led to 2.3$\times$-5.2$\times$ speedups over a blocked version. Bondhugula et al.~\cite{floyd_fpga} proposed a tiled parallel implementation using Field Programmable Gate Arrays. In the field of GPUs, we highlight the work of Katz and Kider~\cite{floyd_katz_gpu}, who proposed a shared memory cache efficient implementation to handle graph sizes that are inherently larger than the DRAM memory available on the device. Also, Matsumoto et al.~\cite{floyd_matsumoto_cpu_gpu} presented a blocked algorithm for hybrid CPU-GPU systems aimed to minimize host-device communication. Finally, Hou et al.~\cite{floyd_knc} evaluated different optimization techniques for Xeon Phi KNC coprocessor. Just as this study, they found that blocking and vectorization are key aspects in this problem to achieve high performance. Also, guided vectorization led to better results than the manual approach, but with larger performance differences. Contrary to this work, their implementation benefited from using more than one thread per core. However, as stated before, there are significant architectural differences between these platforms that support this behavior.

\section{Conclusions}
\label{sec:conclusion}

KNL is the second generation of Xeon Phi family and features new technologies in SIMD execution and memory access.
In this paper, we have evaluated a set of programming and optimization techniques for these processors taking the FW algorithm as a representative case study of graph and memory-bound applications. Among the main contributions of this research we can summarize:
\begin{itemize}
\item Blocking technique not only improved performance but also allowed us to apply a coarse-grain workload distribution in the parallel implementation.

\item SIMD exploitation was crucial to achieve top performance. In particular, the serial version run 2.9$\times$, 6$\times$ and 15.5$\times$ faster using the SSE, AVX2 and AVX-512 extensions, respectively.
\item Aligning memory accesses and loop unrolling also showed significant speedups.
\item A single thread per core was enough to get maximal performance. In addition, \emph{scatter} and \emph{balanced} affinities provided extra performance.
\item Besides keeping portability, guided vectorization led to slightly better performance than the intrinsic counterpart, running upto 1.03$\times$ faster. 
\item MCDRAM usage demonstrated to be an efficient strategy to tolerate high-bandwidth demands with practically null programmer intervention, even when the dataset largely exceeded the MCDRAM size. In particular, it produced an average speedup of 9.8$\times$ and a maximum speedup of 15.5$\times$
\end{itemize}

As future work, we consider evaluating programming and optimization techniques in other cluster and memory modes as a way to extract more performance.

\subsubsection*{Acknowledgments}
The authors thank the ArTeCS Group from Universidad Complutense de Madrid for letting use their Xeon Phi KNL system.

%-------------------------------------------------------------------------------
\bibliographystyle{splncs03}
\bibliography{biblio}

\end{document}